\documentclass[conference]{IEEEtran}
\usepackage{graphicx,amssymb,amsmath}%图片支持
\usepackage{multicol}%正文双栏
\usepackage{CJK}
\usepackage[noadjust]{cite}%参考文献
\usepackage{setspace}%调整行距
\usepackage{stfloats}
\usepackage{midfloat}
\usepackage{flushend,cuted}
\usepackage{bm}
\usepackage{multirow}
\usepackage{array}
\usepackage{stfloats}
\usepackage{cite}
\usepackage{algorithm}% http://ctan.org/pkg/algorithms
\usepackage{algpseudocode}% http://ctan.org/pkg/algorithmicx

\IEEEoverridecommandlockouts

\ifCLASSINFOpdf
\else
\fi
\newcommand{\tr}{\textmd{tr}}
\newcommand{\E}{\textmd{E}}

\newcommand{\diag}{\textmd{diag}}
\newcommand{\vect}{\textmd{vec}}
\newcounter{MYtempeqncnt}

\begin{document}
\title{How Much Training is Needed in One-Bit Massive MIMO Systems at Low SNR?}

\author{\IEEEauthorblockN{Yongzhi Li\IEEEauthorrefmark{1},
Cheng Tao\IEEEauthorrefmark{1},
Liu Liu\IEEEauthorrefmark{1},
Amine Mezghani\IEEEauthorrefmark{2}, and
A. Lee Swindlehurst\IEEEauthorrefmark{2}}
\IEEEauthorblockA{\IEEEauthorrefmark{1} Institute of Broadband Wireless Mobile Communications, Beijing Jiaotong University, Beijing 100044, P.R.China.}
%\IEEEauthorblockA{\IEEEauthorrefmark{2} Universitat Aut$\grave{\textrm{o}}$noma de Barcelona, Bellaterra, Barcelona 08193, Spain.}
\IEEEauthorblockA{\IEEEauthorrefmark{2} Center for Pervasive Communications and Computing (CPCC), University of California, Irvine, Irvine, CA 92697, USA.  \\
(Email: \{liyongzhi, chtao, liuliu\}@bjtu.edu.cn, amezghani@uci.edu, swindle@uci.edu)}
%Telephone: (800) 555--1212, Fax: (888) 555--1212}
%\IEEEauthorblockA{\IEEEauthorrefmark{4}Tyrell Inc., 123 Replicant Street, Los Angeles, California 90210--4321}
}
\maketitle

\begin{abstract}
%\boldmath
This paper considers training-based transmissions in massive multi-input multi-output (MIMO) systems with one-bit analog-to-digital converters (ADCs). We assume that each coherent transmission block consists of a pilot training stage and a data transmission stage. The base station (BS) first employs the linear minimum mean-square-error (LMMSE) method to estimate the channel and then uses the maximum-ratio combining (MRC) receiver to detect the data symbols. We first obtain an approximate closed-form expression for the uplink achievable rate in the low SNR region. Then based on the result, we investigate the optimal training length that maximizes the sum spectral efficiency for two cases: i) The training power and the data transmission power are both optimized; ii) The training power and the data transmission power are equal. Numerical results show that, in contrast to conventional massive MIMO systems, the optimal training length in one-bit massive MIMO systems is greater than the number of users and depends on various parameters such as the coherence interval and the average transmit power. Also, unlike conventional systems, it is observed that in terms of sum spectral efficiency, there is relatively little benefit to separately optimizing the training and data power.
\end{abstract}

%\begin{IEEEkeywords}
%IEEEtran, journal, \LaTeX, paper, template.
%\end{IEEEkeywords}

% \fi
\IEEEpeerreviewmaketitle

\section{Introduction}
Channel state information (CSI) plays a crucial role for high data rate transmission in wireless communications, especially for massive multi-input multi-output (MIMO) systems. It has been shown that with CSI known at the base station (BS), massive MIMO techniques can average out the noise and interference among the terminals, and hence significantly improve the spectral efficiency even when employing simple signal processing techniques such as maximum-ratio combining (MRC) \cite{marzetta2010noncooperative, ngo2013energy}.

However, with a large number of antenna elements deployed at the BS, system cost and power consumption will be excessive if each antenna element and corresponding radio-frequency (RF) chain is equipped with a high-resolution and power-hungry analog-to-digital converter (ADC). In addition, as huge bandwidths and correspondingly high sampling rates will be required in next generation wireless systems, high-speed ADCs are either unavailable or too costly for practical implementation \cite{texas}. Therefore, finding alternative approaches is needed.

One-bit ADCs are of particular interest since they consist of a simple comparator, and hence have the lowest cost and power consumption. In addition, it has been shown in \cite{mezghani2008analysis,nossek2006capacity} that the capacity of MIMO systems is not severely reduced by the coarse quantization and the power penalty due to the one-bit quantization is approximately equal to only $\pi/2$ at low signal-to-noise ratio (SNR). Therefore, one-bit ADCs can potentially make massive MIMO more viable in practice, especially in low SNR scenarios where such systems are likely to operate.

%One possible solution to this problem is to employ one-bit ADCs, which consist of a simple comparator, and hence have the lowest cost and power consumption. Intuitively, the one-bit ADCs would suffer from a severe system performance loss compared to the high-resolution ADCs. However, it has been shown in \cite{mezghani2008analysis,nossek2006capacity} that the capacity of MIMO system is not severely reduced by the coarse quantization and the power penalty due to the one-bit quantization is approximately equal to only $\pi/2$ at low signal-to-noise ration (SNR). Therefore, the one-bit ADCs can make massive MIMO more feasible and viable in practice.

There has been some recent work on one-bit massive MIMO, particularly focused on pilot-based channel estimation \cite{chiara2014massive,juncil2015near,jianhua2014channel,jacobsson2015one,yongzhi2016channel,li2016optimal}, and several different channel estimators have been proposed. In particular, \cite{li2016optimal} investigated the optimal training length for uplink massive MIMO systems with low-resolutions ADCs. However, it employed the additive quantization noise model (AQMN) and only considered the case where training power and data transmission power are the same. In this paper, we evaluate the training duration that optimizes the sum spectral efficiency in one-bit massive MIMO at low SNR by employing Bussgang decomposition. We derive an approximate closed-form expression for the uplink achievable rate with the linear minimum mean-square-error (LMMSE) channel estimate in the low SNR region. Based on the approximation, we focus on the problem of how much of the coherence interval should be spent on training to maximize the sum spectral efficiency for two cases: (i) where the users can employ different power during training and data transmission, and (ii) where the users employ the same training and data transmission power. Numerical results show that the optimal training duration in one-bit massive MIMO system depends on various system parameters. In particular, using the same power for training and data transmission is seen to achieve a sum spectral efficiency close to that in the case where power is optimized, and hence we conclude that using the same power should be preferred since in practice the users often do not have the luxury of varying the power during the training and data transmission stages.

\section{System Model and Channel Estimation}
\subsection{System Model}
We consider a single-cell one-bit massive MIMO system with $K$ single-antenna terminals and an $M$-antenna BS. For uplink data transmission, the received signal at the BS is
\begin{equation}
{\mathbf{y}} = \sqrt {{\rho_d}}\, {\mathbf{Hs}} + {\mathbf{n}},
\end{equation}
where the elements of the channel $\mathbf{H}$ are distributed as $\mbox{\rm vec}(\mathbf{H}) = \mathbf{\underline{h}} \sim \mathcal{CN}(0,\mathbf{I})$ is the $M\times K$ channel matrix, $\mathbf{n}\sim\mathcal{CN}(0,\mathbf{I})\in\mathbb{C}^{M\times 1}$ denotes additive white Gaussian noise, and $\mathbf{s}$ is a vector containing the signal transmitted by each user. We assume $\E\{|s_k|^2\} = 1$ and hence we define the scale factor $\rho_d$ to be the uplink SNR. The quantized signal obtained after the one-bit ADCs is represented as
\begin{equation}\label{sysmtem_quantization}
\mathbf{r} = \mathcal{Q}(\mathbf{y}) = \mathcal{Q}(\sqrt {{\rho_d}}\, {\mathbf{Hs}} + {\mathbf{n}}),
\end{equation}
where $\mathcal{Q}(.)$ represents the one-bit quantization operation, which is applied separately to the real and imaginary parts of the signal. The outcome of the one-bit quantization thus lies in the set $\mathcal{R} = 1/\sqrt 2 \{1+1j, 1-1j, -1+1j, -1-1j\}$.

\subsection{Channel Estimation}
In a practical system, the channel $\mathbf{H}$ has to be estimated at the BS. In the uplink transmission phase, we assume that the channel coherence interval is divided into two parts: one dedicated to training and the other to data transmission.

For the training stage, we assume all users simultaneously transmit pilot sequences of $\tau$ symbols to the BS, which yields
\begin{equation}
\mathbf{Y}_p = \sqrt{\rho_p}\mathbf{H}\bm{\Phi}^T + \mathbf{N}_p,
\end{equation}
where $\mathbf{Y}_p \in\mathbb{C}^{M\times \tau}$ is the received signal, $\rho_p$ is the transmit power of each pilot symbol, and $\bm{\Phi}\in\mathbb{C}^{\tau\times K}$ is the matrix of pilot symbols. Vectorizing the received signal yields
\begin{align}
\mathbf{y}_p &= \vect(\sqrt{\rho_p}\mathbf{H}\bm{\Phi}^T + \mathbf{N}_p) \nonumber \\
& = (\bm{\Phi}\otimes \sqrt{\rho_p}\mathbf{I}_M)\underline{\mathbf{h}}+\mathbf{n}_p =\bar{\bm{\Phi}}\underline{\mathbf{h}} + \mathbf{n}_p,
\end{align}
where $\mathbf{n}_p = \vect(\mathbf{N}_p)$. We can see from~\eqref{sysmtem_quantization} that after the nonlinear operation $\mathcal{Q}(.)$ of the one-bit ADCs, the amplitude information of the the received signal is lost and only the sign information remains. However, using the Bussgang decomposition \cite{bussgang1952yq}, we can reformulate the nonlinear quantization with a statistically equivalent linear operator that will simplify the channel estimator and the resulting analysis. In particular, for the one-bit quantizer in \eqref{sysmtem_quantization}, the Bussgang decomposition is written
\begin{align}\label{r_p}
\mathbf{r}_p &= \mathcal{Q}(\mathbf{y}_p) = \mathbf{A}_p\mathbf{y}_p + \mathbf{q}_p  = \tilde{\bm{\Phi}}\underline{\mathbf{h}} + \tilde{\mathbf{n}}_p,
\end{align}
where the $i$th element of $\mathbf{r}_p$ takes values from the set $\mathcal{R}$, $\tilde{\bm{\Phi}} = \mathbf{A}_p(\bm{\Phi} \otimes \sqrt{\rho_p}\mathbf{I})$, $\tilde{\mathbf{n}}_p = \mathbf{A}_p\mathbf{n}_p + \mathbf{q}_p$, $\mathbf{A}_p$ is the linear operator of the Bussgang decomposition, and $\mathbf{q}_p $ the statistically equivalent quantization noise. The matrix $\mathbf{A}_p$ is chosen to make $\mathbf{q}_p$ uncorrelated with (but still dependent on) $\mathbf{y}_p$ \cite{bussgang1952yq}, or equivalently, to minimize the power of the equivalent quantization noise. For one-bit quantization, we have \cite{yongzhi2016channel}
\begin{align}\label{Optimal_A_p}
\mathbf{A}_p &= \sqrt{\frac{2}{\pi}}\diag(\mathbf{C}_{\mathbf{y}_p\mathbf{y}_p})^{-\frac{1}{2}} \nonumber \\
& = \sqrt {\frac{2}{\pi }} \diag\left(\left(\bm{\Phi}\bm{\Phi}^H \otimes \rho_p\mathbf{I}_{M}\right) + \mathbf{I}_{M\tau}\right)^{-\frac{1}{2}}.
\end{align}

{\it Remark 1:} We can see from \eqref{Optimal_A_p} that $\mathbf{A}_p$ is related to the diagonal terms of $\mathbf{C}_{\mathbf{y}_p\mathbf{y}_p}$ and therefore to the pilot matrix. In order to obtain a simple expression for $\mathbf{A}_p$, in \cite{yongzhi2016channel} random pilot sequences with $\tau =K$ were chosen. In this paper, however, we relax this constraint and allow for the possibility of $\tau\ge K$. In addition, we consider pilot sequences composed of submatrices of the discrete Fourier transform (DFT) operator. The benefits of using DFT pilot sequences are: i) all the elements of the matrix have the same magnitude, which simplifies peak transmit power constraints, and ii) the diagonal terms of $\bm{\Phi}\bm{\Phi}^H$ are always equal to $K$, which results in a simple expression for $\mathbf{A}_p$, as follows:
\begin{equation}
\mathbf{A}_p = \sqrt{\frac{2}{\pi}\frac{1}{K\rho_p + 1}} \mathbf{I} = \alpha_p\mathbf{I}.
\end{equation}

According to \cite{kay1993fundamentals} and the fact that $\mathbf{q}_p$ is uncorrelated with the channel $\underline{\mathbf{h}}$ \cite{yongzhi2016channel}, the LMMSE channel estimate of $\mathbf{\underline{h}}$ can be expressed as
\begin{equation}\label{channel_estimate}
\hat{\underline{\mathbf{h}}} = \tilde{\bm{\Phi}}^H\mathbf{C}_{\mathbf{r}_p\mathbf{r}_p}^{-1}\mathbf{r}_p,
\end{equation}
where $\mathbf{C}_{\mathbf{r}_p\mathbf{r}_p}$ is the auto-correlation matrix of $\mathbf{r}_p$ given by
\begin{align}\label{C_rr}
\mathbf{C}_{\mathbf{r}_p\mathbf{r}_p} =& \frac{2}{\pi} \left(\arcsin\left(\bm{\Sigma}^{-\frac{1}{2}}_{\mathbf{y}_p \mathbf{y}_p} \Re\left({\mathbf{C}_{\mathbf{y}_p \mathbf{y}_p}}\right) \bm{\Sigma}^{-\frac{1}{2}}_{\mathbf{y}_p \mathbf{y}_p} \right) \right. \nonumber \\
& \left. + j \arcsin\left(\bm{\Sigma}^{-\frac{1}{2}}_{\mathbf{y}_p \mathbf{y}_p} \Im\left({\mathbf{C}_{\mathbf{y}_p \mathbf{y}_p}}\right) \bm{\Sigma}^{-\frac{1}{2}}_{\mathbf{y}_p \mathbf{y}_p} \right)  \right).
\end{align}
The normalized MSE of the BLMMSE channel estimate is thus
\begin{align}
\textrm{MSE} &=\frac{1}{MK}{\E\left\{\left\|\underline{\mathbf{h}} - \hat{\underline{\mathbf{h}}}\right\|_2^2\right\}}\nonumber \\
\label{nMSE2}& = \frac{1}{MK}\tr\left(\mathbf{I} - \tilde{\bm{\Phi}}^H\mathbf{C}_{\mathbf{r}_p\mathbf{r}_p}^{-1}\tilde{\bm{\Phi}} \right).
\end{align}

{\it Remark 2:} Each element of $\hat{\mathbf{h}}$ can be expressed as a summation of a large number of random variables, i.e., $[\hat{\mathbf{h}}]_{n} = \sum_{i=1}^{M\tau}[\tilde{\bm{\Phi}}^H\mathbf{C}_{\mathbf{r}_p\mathbf{r}_p}]_{n,i} r_{p,i}$. Although the elements of the channel estimate \eqref{channel_estimate} are not exactly Gaussian distribution due to the one-bit quantization, we can approximate it as Gaussian according to Cram{\' e}r's central limit theorem \cite{Cramer2004random}. Therefore, in the sequel we model each element of the channel estimate $\hat{\mathbf{h}}$ as Gaussian with zero mean and variance $\eta^2 = \tr\left(\tilde{\bm{\Phi}}^H\mathbf{C}_{\mathbf{r}_p\mathbf{r}_p}^{-1}\tilde{\bm{\Phi}}\right)/MK$.
%We can see from \eqref{C_rr} that, the expression of the auto-correlation matrix of $\mathbf{C}_{\mathbf{r}_p\mathbf{r}_p}$ is complicated since it involves the arcsin operation. Therefore, it is difficult to express the MSE in a simple expression. However, the arcsin operation can be avoided in the low SNR region by using the approximation we provide in the next section.

\section{Uplink Achievable Rate Analysis}
In the data transmission stage, we assume $K$ users simultaneously transmit their data symbols, represented as $\mathbf{s}$, to the BS. After one-bit quantization, the signal at the BS can be expressed as
\begin{align}
\mathbf{r}_d &= \mathcal{Q}(\mathbf{y}_d) = \mathcal{Q}(\sqrt{\rho_d}\mathbf{Hs} + \mathbf{n}_d) \nonumber \\
& = \sqrt{\rho_d}\mathbf{A}_d \mathbf{Hs} + \mathbf{A}_d\mathbf{n}_d + \mathbf{q}_d,
\end{align}
where the same definitions as in previous sections apply, but with the subscript $p$ replaced with $d$. Following the same reasoning as in Section II.B, in order to minimize the quantization noise (or equivalently, to make it uncorrelated with $\mathbf{y}_d$), we can use the Bussgang decomposition to represent the model with $\mathbf{A}_d = \alpha_d\mathbf{I}$ and $\alpha_d = \sqrt{2/(\pi(1+K\rho_d))}\mathbf{I}$.

Next, we assume that the BS regards the LMMSE channel estimate as the true channel and employs the MRC receiver to detect the data symbols transmitted by the $K$ users. For the MRC receiver, the quantized signal is separated into $K$ streams by multiplying it with $\hat{\mathbf{{H}}} = \vect^{-1}(\hat{\underline{\mathbf{h}}})$:
\begin{align}\label{MRC}
\hat{\mathbf{s}} &= \hat{\mathbf{H}}^H \mathbf{r}_d \nonumber \\
& = \sqrt{\rho_d}\hat{\mathbf{H}}^H \mathbf{A}_d(\hat{\mathbf{H}}\mathbf{s} + \bm{\mathcal{E}}\mathbf{s}) + \hat{\mathbf{H}}^H\mathbf{A}_d\mathbf{n}_d + \hat{\mathbf{H}}^H\mathbf{q}_d,
\end{align}
where $\bm{\mathcal{E}} = \mathbf{H}-\hat{\mathbf{H}}$ denotes the channel estimation error. As such, the $k$th element of $\hat{\mathbf{s}}$ is used to decode the signal transmitted from the $k$th user:
\begin{align}\label{hat_s_k}
{\hat s_k} =& {\sqrt {{\rho _d}} \hat{\mathbf{{h}}}_k^H{{\bf{A}}_d}{\hat{\bf{h}}_k}{s_k}}+  {\sqrt {{\rho _d}} \hat{\mathbf{{h}}}_k^H \sum \nolimits_{i \ne k}^K {{\bf{A}}_d}{\hat{\bf{h}}_i}{s_i}} \nonumber\\
  &+{\sqrt {{\rho _d}} \hat{\mathbf{{h}}}_k^H \sum \nolimits_{i =1}^K {{\bf{A}}_d}\bm{\varepsilon}_i{s_i}} + {\hat{\mathbf{{h}}}_k^H{{\bf{A}}_d}{\bf{n}}_d} + {\hat{\mathbf{{h}}}_k^H{\bf{q}}_d},
\end{align}
where $\hat{\mathbf{{h}}}_i$ and $\bm{\varepsilon}_i$ are the $i$th column of $\hat{\mathbf{{H}}}$ and $\bm{\mathcal{E}}$, respectively. The last four terms in \eqref{hat_s_k} correspond respectively to user interference, channel estimation error, AWGN noise and the quantization noise.

Note that although $\mathbf{q}_d$ is not Gaussian due to the one-bit quantization, the worst-case additive noise that minimizes the input-output mutual information is Gaussian \cite{hassibi2003how}, and hence a lower bound for the achievable rate can be found by modeling $\mathbf{q}_d$ as Gaussian with the same covariance matrix:
\begin{equation}
\mathbf{C}_{\mathbf{q}_d\mathbf{q}_d} = \mathbf{C}_{\mathbf{r}_d\mathbf{r}_d} -\mathbf{A}_d\mathbf{C}_{\mathbf{y}_d\mathbf{y}_d}\mathbf{A}_d^H.
\end{equation}
Thus, the ergodic achievable rate of the uplink transmission in one-bit massive MIMO is lower bounded by \eqref{ergodic_achievable_rate}, shown on the top of next page. Since there is no efficient way to directly calculate the achievable rate in \eqref{ergodic_achievable_rate}, we provide an approximation in the following theorem:

\begin{figure*}[!t]
\normalsize
\setcounter{MYtempeqncnt}{\value{equation}}
\setcounter{equation}{14}
\begin{equation}\label{ergodic_achievable_rate}
\footnotesize
\tilde{R}_k = \E\left\{\log_2\left(1+\frac{\rho_d \left|\hat{\mathbf{{h}}}_k^H{{\bf{A}}_d}{\hat{\bf{h}}_k}\right|^2}{\rho_d\sum_{i \ne k}^K\left|\hat{\mathbf{{h}}}_k^H{{\bf{A}}_d}{\hat{\bf{h}}_i}\right|^2 + \rho_d\sum_{i =1}^K\left|\hat{\mathbf{{h}}}_k^H{{\bf{A}}_d}{\bm{\varepsilon}_i}\right|^2+ \left\|\hat{\mathbf{{h}}}_k^H{{\bf{A}}_d}\right\|^2 + \hat{\mathbf{{h}}}_k^H \mathbf{C}_{\mathbf{q}_d\mathbf{q}_d}\hat{\mathbf{{h}}}_k }\right)\right\}
\end{equation}
\setcounter{equation}{15}
\hrulefill
\vspace*{-0.3cm}
\end{figure*}

{\it Theorem 1}: For an MRC receiver based on the LMMSE channel estimate, the uplink achievable rate of the $k$th user in a
one-bit massive MIMO system can be approximated by
\begin{equation}\label{closed_form}
R_k = \log_2\left(1+\frac{\rho_d\alpha_d^2\eta^2(M+1)}{\rho_d\alpha_d^2(K-\eta^2)+\alpha_d^2+1-2/\pi}\right),
\end{equation}
where $\eta^2 = \tr\left(\tilde{\bm{\Phi}}^H\mathbf{C}_{\mathbf{r}_p\mathbf{r}_p}^{-1}\tilde{\bm{\Phi}}\right)/MK$.
\begin{IEEEproof}
See Appendix A.
\end{IEEEproof}

\section{Optimal Training Length in Low SNR Region}
Although \cite{li2016optimal} investigated the optimal training length for uplink massive MIMO systems with low-resolutions ADCs. However, it employed the AQMN and only considered the case where training power and data transmission power are the same. In the analysis below, we first derive the approximation of sum spectral efficiency at low SNR. Based on the approximation, we then evaluate the optimal training length that maximizes the sum spectral efficiency considering two case: i) the training power and data transmission power are both optimized; ii) the training power and data transmission power are the same.

\subsection{Low SNR Sum Spectral Efficiency Approximation}
We first define the sum spectral efficiency as the sum rate per channel use. Let $T$ be the length of the coherence interval in symbols. During each coherence interval, $\tau$ symbols are used for pilot training and the remaining $T-\tau$ symbols are used for data transmission. Therefore, the sum spectral efficiency is given by
\begin{equation}
S = \frac{T-\tau}{T} \sum_{k=1}^K {R}_k.
\end{equation}

We can see that the closed-form expression for the achievable rate in Theorem~1 involves the auto-correlation matrix of $\mathbf{C}_{\mathbf{r}_p\mathbf{r}_p}$, which, according to \eqref{C_rr}, is complicated due to the arcsin operation. However, it is expected that massive MIMO systems will operate at low SNR due to the availability of a large array gain. In what follows, we show that using a low SNR assumption allows us to derive an approximation for $\mathbf{C}_{\mathbf{r}_p\mathbf{r}_p}$ to avoid the arcsin operation in the low SNR region.

According to \eqref{r_p}, we can rewrite the auto-correlation matrix $\mathbf{C}_{\mathbf{r}_p\mathbf{r}_p}$ as
\begin{equation}\label{C_rr_app1}
\mathbf{C}_{\mathbf{r}_p\mathbf{r}_p} = \tilde{\bm{\Phi}}\tilde{\bm{\Phi}}^H+\mathbf{A}_p\mathbf{A}_p^H +
\mathbf{C}_{\mathbf{q}_p\mathbf{q}_p},
\end{equation}
where
\begin{align}\label{C_qq}
\mathbf{C}_{\mathbf{q}_p\mathbf{q}_p} &= \mathbf{C}_{\mathbf{r}_p\mathbf{r}_p} - \mathbf{A}_p\mathbf{C}_{\mathbf{y}_p\mathbf{y}_p}\mathbf{A}_p^H \nonumber \\
& = \frac{2}{\pi}(\arcsin(\mathbf{X})+j\arcsin(\mathbf{Y})) - \frac{2}{\pi}(\mathbf{X} + j\mathbf{Y}),
\end{align}
and where we define
\begin{align}
\mathbf{X} &= \bm{\Sigma}^{-\frac{1}{2}}_{\mathbf{y}_p \mathbf{y}_p} \Re\left({\mathbf{C}_{\mathbf{y}_p \mathbf{y}_p}}\right)\bm{\Sigma}^{-\frac{1}{2}}_{\mathbf{y}_p \mathbf{y}_p} \\
\mathbf{Y} &= \bm{\Sigma}^{-\frac{1}{2}}_{\mathbf{y}_p \mathbf{y}_p} \Im\left({\mathbf{C}_{\mathbf{y}_p \mathbf{y}_p}}\right)\bm{\Sigma}^{-\frac{1}{2}}_{\mathbf{y}_p \mathbf{y}_p}.
\end{align}

Note that the ``arcsin'' is an element-wise operation, and it can be approximated as
\begin{equation}
\frac{2}{\pi}\arcsin(a) \cong \left\{ \begin{array}{*{20}{c}}
{1,}&{a = 1}\\
{2a/\pi,}&{a<1}
\end{array}
\right.
\end{equation}
Since the non-diagonal elements of $\mathbf{X}$ and $\mathbf{Y}$ are far smaller than 1 in the low SNR region, we can approximate \eqref{C_qq} as
\begin{equation}\label{C_qq_app}
\mathbf{C}_{\mathbf{q}_p\mathbf{q}_p} \cong (1-2/\pi)\mathbf{I}.
\end{equation}
Substituting \eqref{C_qq_app} and \eqref{C_rr_app1} into the expression for $\eta^2$, we have
\begin{align}
\eta^2 &\cong \tr\left(\tilde{\bm{\Phi}}^H(\tilde{\bm{\Phi}}\tilde{\bm{\Phi}}^H+(\alpha_p^2+1-2/\pi)\mathbf{I})^{-1}\tilde{\bm{\Phi}}\right)/MK \nonumber \\
& = (\alpha_p^2 \tau\rho_p + \alpha_p^2+1-2/\pi)^{-1}\alpha_p^2 \tau\rho_p = \sigma^2.
\end{align}
The equation on the second line holds due to the matrix inversion identity $(\mathbf{I} + \mathbf{AB})^{-1}\mathbf{A} = \mathbf{A}(\mathbf{I} + \mathbf{BA})^{-1}$. Therefore in the low SNR region, we can approximate  the sum spectral efficiency as
\begin{equation}\label{SE_low}
\footnotesize
S^{\textrm{low}} = \frac{(T-\tau)K}{T} \log_2\left(1+\frac{\rho_d\alpha_d^2\sigma^2(M+1)}{\rho_d\alpha_d^2(K-\sigma^2)+\alpha_d^2+1-2/\pi}\right).
\end{equation}

\subsection{Optimal Training Duration for One-Bit Massive MIMO Systems}

Let $\rho$ be the average transmit power and $P = \rho T$ be the total energy budget for each user over the coherence interval, which satisfies the constraint $\tau\rho_p + (T-\tau)\rho_d\le P$. For any power allocation in which some users do not expend their full energy budget, such users could increase their training power to improve their own achievable rate without causing interference to other users. Thus, we can replace the inequality constraint on the total energy budget with the equality constraint $\tau\rho_p + (T-\tau)\rho_d = P$. Thus, the optimization problem can be expressed as
\begin{align}
& {\text{maximize}}
& & S^{\textrm{low}} \nonumber\\ % ,\rho_d = \frac{(1-\lambda) P}{T-K}}
& \text{subject to} & & \tau\rho_p + (T-\tau)\rho_d = P, \nonumber \\
\label{opt_general}& & & K\le\tau\le T.
\end{align}

Next we focus on the optimal training duration problem and consider two cases: (i) The training power and data transmission power are both optimized; (ii) The training power is equal to the data transmission, $\rho_p = \rho_d = \rho$. The latter case is of interest since the users may not have the ability to change their transmit power from the training to the data transmission phases.

{\it Case I}: For the first case, we assume the users can vary the training power and the data transmission power and jointly choose $\{\tau,\rho_p,\rho_d\}$ to maximize the sum spectral efficiency. To facilitate the presentation, let $\gamma \in(0,1)$ denote the fraction of the total energy budget that is devoted to pilot training, such that $\gamma P = \tau\rho_p$ and $(1-\gamma)P = (T-\tau)\rho_d$.
Thus the optimization problem of \eqref{opt_general} can be rewritten as
\begin{align}
& {\text{maximize}}
& & S^{\textrm{low}}|_{\rho_p = \frac{\gamma P}{\tau},\rho_d = \frac{(1-\gamma) P}{T-\tau}}, \nonumber\\ % ,\rho_d = \frac{(1-\lambda) P}{T-K}}
\label{opt_I} & \text{subject to} & & 0<\gamma<1,~K\le\tau\le T.
\end{align}

Note that previous work \cite{hassibi2003how} has shown that for conventional MIMO systems with infinite precision ADCs, the optimal training duration is always $\tau^* = K$. However, we will see that this is not the case for one-bit massive MIMO. First we rewrite the sum spectral efficiency of \eqref{SE_low} as a function with respect to $\gamma$ and $\tau$:
\begin{equation}\label{SE_r_t}
\footnotesize
S^{\textrm{low}}(\gamma,{\tau}) = \frac{(T-\tau)K}{T}\log_2\left(1+\frac{a_1\tau}{a_2\tau^2+a_3\tau+a_4}\right),
\end{equation}
where we define
\begin{align*}
a_1 =& 4(M+1)(\gamma-\gamma^2)P^2,~~~a_2  = (\pi^2 + 2 P \pi \gamma),  \\
a_3=& 4 P^2 (-1 + \gamma) \gamma +  K P \pi (\pi - 2 \pi \gamma + 2 \gamma (1 + P - P \gamma)) \\
&+ \pi^2 T +  2 P \pi \gamma T , \\
a_4=& K^2 P^2 ( \pi^2 -2\pi) (\gamma - \gamma^2) + K P ( \pi^2 -2\pi ) \gamma T .
\end{align*}
Then we denote $\{\gamma^*,\tau^*\}$ to be the solution of \eqref{opt_I}, such that $\gamma^* P = \tau^*\rho_p^*$ is the optimal energy for training, and $(1-\gamma^*) P = (T-\tau^*)\rho_d^*$ is the optimal amount for data transmission. Next we choose $\bar{\tau} = K$, $\bar{\rho}_p = \gamma^*P/\bar{\tau} $ and $\bar{\rho}_d = (1-\gamma^*)P/(T-\bar{\tau})$. Clearly, the function of \eqref{SE_r_t} is not a monotonic function with respect to $\tau$ with a given $\gamma^*$. That is to say, it is difficult to compare the values of $S^{\textrm{low}}(\gamma^*,\tau^*)$ and $S^{\textrm{low}}(\gamma^*,\bar{\tau})$. Although we cannot obtain a closed-form expression for $\tau^*$, we can numerically determine $\tau^*$ and $\gamma^*$. For the simulations in the next section, we used the \verb"fmincon" function in Matlab for the optimization. As we will show in the next section, unlike conventional MIMO systems, the optimal training duration depends on the coherence interval $T$ and the total energy budget $P$.

{\it Case II}: In this case, the optimization problem of \eqref{opt_general} simplifies to
\begin{align}
& {\text{maximize}}
& & S^{\textrm{low}}|_{\rho_p =\rho_d = \rho}, \nonumber\\
\label{opt_II} & \text{subject to} & & K\le\tau\le T.
\end{align}
Obviously, there exists a tradeoff between the training duration $\tau$ and the data transmission duration $T-\tau$. As we increase $\tau$, the accuracy of the channel estimate improves, thereby increasing the sum spectral efficiency. On the other hand, as $\tau$ increases, the data transmission duration decreases, thereby decreasing the sum spectral efficiency. As in the previous case, we obtain the optimal $\tau$ by solving~\eqref{opt_II} numerically.

\section{Numerical Results}

For the simulations, we assume a one-bit massive MIMO system with $M = 128$ BS antennas and $K=8$ users. In all plots, the curves for conventional massive MIMO are obtained using the approximate closed-form expression of the uplink achievable rate from \cite{ngo2013energy}.

\begin{figure}
  \centering
  \includegraphics[width=7cm]{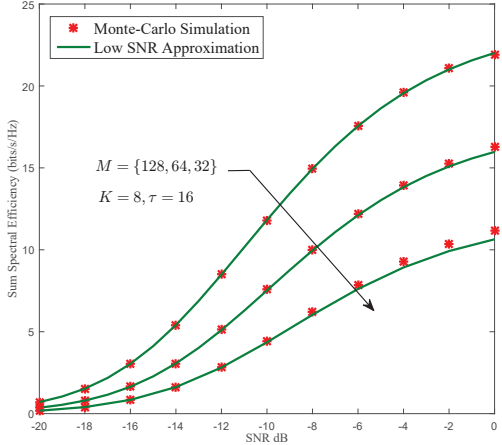}\\
  \vspace{-0.5cm}
  \caption{Sum spectral efficiency versus SNR with $M=\{32,64,128\}$, $K=8$, $T = 200$, $\tau = 16$.}\label{SE_MonteCarlo_App}
  \vspace{-0.4cm}
\end{figure}
We first evaluate the validity of our obtained approximate expression of the achievable rate with the ergodic expressions given in~\eqref{ergodic_achievable_rate} and~\eqref{SE_low}, respectively. Figure~\ref{SE_MonteCarlo_App} illustrates the sum spectral efficiency versus SNR with different numbers of transmit antennas $M=\{32, 64, 128\}$ for $T = 200$, and $\tau = 16$. The results show that the gap between the approximate expression and the ergodic achievable rate can be neglected, and thus in the following plots we use the approximation.

\begin{figure}
  \centering
  \includegraphics[width=7cm]{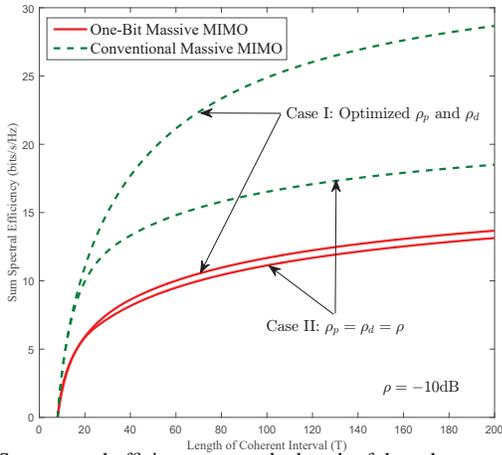}\\
  \vspace{-0.5cm}
  \caption{Sum spectral efficiency versus the length of the coherence interval $T$ for conventional and one-bit massive MIMO systems with $M=128$, $K=8$ and $\rho = -10$dB.}\label{SE_T_TwoCases}
  \vspace{-0.4cm}
\end{figure}
Figure~\ref{SE_T_TwoCases} shows the sum spectral efficiency versus the length of the coherence interval for conventional and one-bit massive MIMO systems with $\rho = -10$dB. We see that the performance gap between the case of optimized $\rho_p$ and $\rho_d$ and the case of $\rho_p = \rho_d = \rho$ is large for conventional massive MIMO system, but almost negligible for one-bit massive MIMO systems. One may conclude from this that the power optimization is not useful for one-bit systems since allowing different power levels between training and data transmission may be a complicated feature to implement at the user terminals.

\begin{figure}[!t]
  \centering
  \includegraphics[width=7cm]{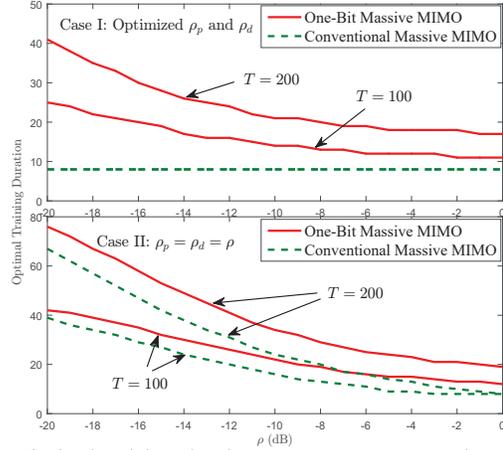}\\
  \vspace{-0.5cm}
  \caption{Optimal training duration versus average transmit power $\rho$ for conventional and one-bit massive MIMO systems with $M=128$, $K=8$ and $T=\{100,200\}$.}\label{OptimalTraining_SNR}
  \vspace{-0.4cm}
\end{figure}

Figure~\ref{OptimalTraining_SNR} compares the optimal training duration versus the average transmit power for conventional and one-bit massive MIMO systems assuming $T = \{100, 200\}$. For conventional massive MIMO systems, the optimal training duration is $\tau^* = K$ for Case I, while it changes with the total energy budget for Case II. However, for one-bit massive MIMO systems, the optimal training duration changes with transmit power in both cases. In addition, sum spectral efficiency is enhanced with more training in one-bit massive MIMO compared with conventional systems for all power levels, indicating that more training is necessary to combat the quantization noise.

\begin{figure}[!t]
  \centering
  \includegraphics[width=7cm]{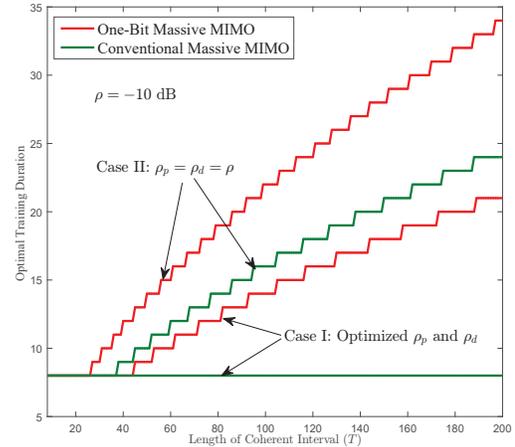}\\
  \vspace{-0.5cm}
  \caption{Optimal training duration versus the length of coherent interval for two cases in conventional massive MIMO and one-bit massive MIMO systems with $M=128$, $K=8$ and $\rho = -10$dB.}\label{OptimalTraining_T}
  \vspace{-0.4cm}
\end{figure}

Figure~\ref{OptimalTraining_T} shows the optimal training duration versus the length of coherence interval with $\rho=-10$dB for conventional and one-bit massive MIMO systems. We again see that, for Case I, the optimal training duration in conventional massive MIMO systems always equals $K$, while in one-bit massive MIMO it increases with $T$. We also see again that the one-bit system requires a larger proportion of the coherence interval devoted to training than in a conventional system.

\section{Conclusions}
This paper has investigated the optimal training duration and training vs. data power allocation that maximizes the sum spectral efficiency for massive MIMO systems with one-bit ADCs. Assuming the BS employs LMMSE channel estimation and the MRC receiver to detect the data symbols, we first obtained an approximate expression for the uplink achievable rate in the low SNR region. Then we optimized this expression over the amount of the coherence interval spent on training for two different power allocations: optimized training and data transmission power, and equal training and data transmission power. When the power allocation is optimized, conventional systems always choose the number of training symbols equal to the number of users, while for one-bit systems the optimal training duration depends on both the coherence interval and the power budget. For equal power allocation, the optimal training duration also varies with these parameters, but one-bit systems always appear to require a higher fraction of symbols devoted to training in order to maximize the sum spectral efficiency.

\section*{Acknowledgment}
This work was supported in part by Beijing Nova Programme (Grant No.xx2016023), Fundamental Research Funds for the Central Universities under grant 2015JBM011, National Natural Science Foundation of China (Grant No. 61471027), Research Fund of National Mobile Communications Research Laboratory, Southeast University (Grant No. 2014D05), and Beijing Natural Science Foundation Project (Grant No. 4152043). A.~Swindlehurst was supported by the National Science Foundation under Grant ECCS-1547155, and by the Technische Universit\"at M\"unchen Institute for Advanced Study, funded by the German Excellence Initiative and the European Union Seventh Framework Programme under grant agreement No. 291763, and by the European Union under the Marie Curie COFUND Program.

\appendices
\section{}\label{proof_Theorem1}
According to \cite[Lemma 1]{zhang2014power}, we can approximate the ergodic achievable rate $\tilde{R}_k$ by
\begin{equation}\label{ergodic_achievable_rate_App}
\footnotesize
R_k = \log_2\left(1+\frac{\rho_d\E\left\{ \left|\hat{\mathbf{{h}}}_k^H{{\bf{A}}_d}{\hat{\bf{h}}_k}\right|^2\right\}}
{\rho_d\E\left\{\left|\hat{\mathbf{{h}}}_k^H{{\bf{A}}_d}{\bm{\varepsilon}_k}\right|^2\right\} +\textrm{UI}_k + \textrm{AN}_k+\textrm{QN}_k}\right) \; ,
\end{equation}
where we define
\begin{equation}
\footnotesize
\textrm{UI}_k = \rho_d\sum\nolimits_{i \ne k}^K\E\left\{\left|\hat{\mathbf{{h}}}_k^H{{\bf{A}}_d}{{\bf{h}}_i}\right|^2\right\}
\end{equation}
\begin{equation}
\footnotesize
\textrm{AN}_k = \E\left\{\left\|\hat{\mathbf{{h}}}_k^H{{\bf{A}}_d}\right\|^2\right\}, ~
\textrm{QN}_k = \E\left\{\hat{\mathbf{{h}}}_k^H \mathbf{C}_{\mathbf{q}_d\mathbf{q}_d}\hat{\mathbf{{h}}}_k\right\} \; ,
\end{equation}
and where the expectation is taken with respect to the channel realizations. For different channel realizations, the covariance matrix of the quantization noise $\mathbf{q}_d$ is given by
\begin{align}
\footnotesize
\E\{\mathbf{q}_d\mathbf{q}_d^H\} &= \E\{\mathbf{r}_d\mathbf{r}_d^H\} - \alpha_d^2\E\{\mathbf{y}_d\mathbf{y}_d^H\}  = (1-2/\pi)\mathbf{I} \; .
\end{align}
By choosing $\mathbf{A}_d = \alpha_d\mathbf{I}$ according to the Bussgang decomposition, $\mathbf{q}_d$ is not only uncorrelated with the received signal $\mathbf{y}_d$, but it is also uncorrelated with the channel $\mathbf{H}$. Therefore, we have
\begin{equation}
\footnotesize
\E\left\{\hat{\mathbf{{h}}}_k^H \mathbf{C}_{\mathbf{q}_d\mathbf{q}_d}\hat{\mathbf{{h}}}_k\right\} = (1-2/\pi) \E\left\{\|\hat{\mathbf{{h}}}_k\|^2\right\} \; .
\end{equation}

Next we calculate the expectation terms shown above. Recall that we model each element of the channel estimate $\hat{\mathbf{\underline{h}}}$ as Gaussian with zero mean and variance $\eta^2$. Hence, each element of the channel estimation error $\mathbf{\mathcal{E}}$ can be modeled as Gaussian with zero mean and variance $1-\eta^2$. Therefore, according to the law of large numbers, we can obtain
\begin{equation}\label{E_1}
\footnotesize
\E\left\{\|\hat{\mathbf{{h}}}_k\|^2\right\} = \eta^2M; ~\E\left\{\left|\hat{\mathbf{{h}}}_k^H{{\bf{h}}_i}\right|^2\right\} = \eta^2 M, ~i\neq k
\end{equation}
\begin{equation}\label{E_2}
\footnotesize
\E\left\{\left|\hat{\mathbf{{h}}}_k^H{{\bf{A}}_d}{\bm{\varepsilon}_k}\right|^2\right\} \cong \alpha_d^2 \eta^2(1-\eta^2)M
\end{equation}
\begin{equation}\label{E_3}
\footnotesize
\E\left\{\left|\hat{\mathbf{{h}}}_k^H{{\bf{A}}_d}{\hat{\bf{h}}_k}\right|^2\right\} = \alpha_d^2 \eta^4(M^2 + M) \; .
\end{equation}
Substituting \eqref{E_1}-\eqref{E_3} into \eqref{ergodic_achievable_rate_App}, we arrive at the result of Theorem 1.
\bibliographystyle{IEEEtran}
\bibliography{reference}

\end{document}